\documentclass[aps,preprint,showpacs,showkeys,nofootinbib,superscriptaddress,tightenlines]{revtex4}

\usepackage{graphicx}  %
\usepackage{amsmath}
\usepackage{bm}  %

\newcommand{\bea}{\begin{eqnarray}}
\newcommand{\eea}{\end{eqnarray}}
\newcommand{\beq}{\begin{equation}}
\newcommand{\eeq}{\end{equation}}
\newcommand{\bqa}{\begin{eqnarray}}
\newcommand{\eqa}{\end{eqnarray}}

\def\mqo2{{\!\!\!}}

\begin{document}

\preprint{INT-PUB-10-019}
\preprint{HISKP-TH-10-12}

\title{~\\
Scattering of an Ultrasoft Pion and the $\bm{X(3872)}$}
\author{Eric Braaten}
\affiliation{Department of Physics,
         The Ohio State University, Columbus, OH\ 43210, USA\\}
\affiliation{Helmholtz-Institut f\"ur Strahlen- und Kernphysik
	and Bethe Center for Theoretical Physics,
	Universit\"at Bonn, 53115 Bonn, Germany\\}
\author{H.-W. Hammer}
\affiliation{Helmholtz-Institut f\"ur Strahlen- und Kernphysik
	and Bethe Center for Theoretical Physics,
	Universit\"at Bonn, 53115 Bonn, Germany\\}
\author{Thomas Mehen}
\affiliation{Physics Department, Duke University, Durham, NC\ 27708, USA\\}
\date{May 10, 2010}

\begin{abstract}
The identification of the $X(3872)$ as a loosely-bound 
charm-meson molecule allows it to be described by an effective 
field theory, called XEFT,
for the $D^* \bar D$, $D \bar D^*$ and $D \bar D \pi$ sector of QCD
at energies small compared to the pion mass.  We point out that this
effective field theory can be extended to the sector 
that includes an additional pion and used
to calculate cross sections for the 
scattering of a pion and the $X(3872)$.
If the collision energy is much smaller than the pion mass, 
the cross sections are completely calculable at leading order 
in terms of the masses and widths of the charm mesons, pion masses,
and the binding energy of the $X(3872)$.
We carry out an explicit calculation of the cross section 
for the breakup of the $X(3872)$ 
into $D^{*+} \bar D^{*0}$ by the scattering of a very low energy $\pi^+$.
\end{abstract}

\smallskip
\maketitle

\section{Introduction}
\label{sec:intro}

The $X(3872)$, which was discovered by the Belle Collaboration 
in 2003 \cite{Choi:2003ue}, is a truly remarkable hadron.
There is increasingly compelling experimental and theoretical evidence
that it is a loosely-bound charm-meson molecule whose particle content is
\beq
X = \frac{1}{\sqrt2}
\left( D^{*0} \bar D^0 + D^0 \bar D^{*0} \right)
\label{X-DD}
\eeq
and whose constituents have a mean separation much larger than normal hadrons.  
Its $J^{PC}$ quantum numbers are $1^{++}$.
Its binding energy, $E_X$, relative to the $D^{*0} \bar D^0$ threshold 
is extremely small, less than 1~MeV.
As a consequence, it has universal properties that are determined by 
$E_X$ or, equivalently, by the large $D^{*0} \bar D^0$ scattering length
in the even charge-conjugation channel \cite{Braaten:2004rn}.

Fleming, Kusunoki, Mehen, and van Kolck have constructed an 
effective field theory called XEFT that can be used to calculate 
corrections to the universal properties of the $X(3872)$
systematically \cite{Fleming:2007rp}.
In this effective field theory, 
the elementary constituents are the neutral charm mesons
$D^0$, $D^{*0}$, $\bar D^0$, and $\bar D^{*0}$ and the neutral pion $\pi^0$.
There are two types of interactions: contact interactions between the 
pairs of charm mesons $D^{*0} \bar D^0$ and $D^0 \bar D^{*0}$
and the pion transitions $D^{*0} \leftrightarrow D^0 \pi^0$
and $\bar D^{*0} \leftrightarrow \bar D^0 \pi^0$.
XEFT was designed to describe systems consisting of $D^{*0} \bar D^0$, 
$D^0 \bar D^{*0}$, and $D^0 \bar D^0 \pi^0$ with total energy
very close to the $D^{*0} \bar D^0$ threshold, including the $X(3872)$.
In Ref.~\cite{Fleming:2007rp}, the authors presented power-counting 
arguments that guarantee that the pion transitions can be treated 
perturbatively.  They calculated the decay rate of the $X(3872)$ 
into $D^0 \bar D^0 \pi^0$ to next-to-leading order in XEFT.
Fleming and Mehen have applied XEFT at leading order to decays of the 
$X(3872)$ into the P-wave charmonium states $\chi_{cJ}$ 
and one or two pions \cite{Fleming:2008yn}.
They factored the amplitude for the decay into a long-distance 
XEFT matrix element and a short-distance factor that can be calculated using
heavy-hadron chiral perturbation theory (HH$\chi$PT).

The original formulation of XEFT has a rather limited domain of validity.  
One limitation is that it does not describe charged charm mesons.
This limits its domain of applicability to the energy region within
a few MeV of the $D^{*0} \bar D^0$ threshold, because the 
$D^{*+} D^-$ threshold is higher only by about 8~MeV.  This limitation 
is easily removed by generalizing XEFT to include the charged charm mesons
$D^+$, $D^{*+}$, $D^-$, and $D^{*-}$ and the charged pions
$\pi^+$ and $\pi^-$.  This generalization
extends the domain of applicability of XEFT to all energies
relative to the $D^{*0} \bar D^0$ threshold that are small compared to 
the $D^*-D$ mass difference, which is approximately equal to
the pion mass $m_\pi \approx 135$~MeV.  
If the P-wave charmonium state $\chi_{c1}(2P)$,
whose quantum numbers are also $1^{++}$, lies in this region,
it may be necessary to also include it as an explicit degree of freedom
in order to carry out accurate calculations beyond leading order.

Canham, Hammer, and Springer have obtained universal results for 
scattering of charm mesons with the $X(3872)$ that depend only on 
the $D^{*0} \bar D^0$ scattering length \cite{Canham:2009zq}.
They calculated the S-wave phase shifts for $D^0 X$ scattering 
and for $D^{*0} X$ scattering by solving the three-body problem 
for the charmed mesons.
The $D^0 X$ scattering length and the $D^{*0} X$ 
scattering length are both about an order of magnitude larger 
than the $D^{*0} \bar D^0$ scattering length.
These universal results can be regarded as the predictions 
of XEFT at zeroth order in the pion transitions.
Thus XEFT can be applied to systems consisting of 
three charm mesons with energy close to the appropriate threshold.

In this paper, we point out that XEFT can also be applied to systems 
consisting of $D^* \bar D^*$, $D^* \bar D \pi$, $D \bar D^{*} \pi$, 
and $D \bar D \pi \pi$ with total energy
close to the $D^* \bar D^*$ threshold.
The scattering processes whose cross sections are calculable
using XEFT include $\pi X$ elastic scattering and 
$\pi X \to D^* \bar D^*$ at collision energy much smaller than $m_\pi$.
In Section~\ref{sec:X3872}, 
we summarize the case for the $X(3872)$ as a loosely-bound 
charm-meson molecule and we describe some of its universal properties.
In Section~\ref{sec:breakup}, 
we calculate the cross section for $\pi^+ X \to D^{*+} \bar D^{*0}$.
In Section~\ref{sec:elastic}, 
we discuss elastic $\pi^+ X$ scattering.
Our results are summarized in Section~\ref{sec:summary}.

\section{The $\bm{X(3872)}$}
\label{sec:X3872}

We begin by summarizing the case for the $X(3872)$ as a loosely-bound 
charm-meson molecule whose particle content is given in Eq.~(\ref{X-DD}).
The only experimental information that is necessary to make this 
identification is the determination of its quantum numbers 
and the measurements of its mass.
The quantum numbers of the $X(3872)$ are $1^{++}$, which follows from
\begin{itemize}
\item
the observation of its decay into $J/\psi \, \gamma$, which implies 
that it is even under charge conjugation \cite{Jpsigamma},
\item
analyses of the momentum distributions from its decay into 
$J/\psi \, \pi^+ \pi^-$, which imply that its spin and parity are 
$1^+$ or $2^-$ \cite{JP},
\item
either the observation of its decays into $D^0 \bar D^0 \pi^0$ \cite{Gokhroo:2006bt},
which disfavors $2^-$ because of angular momentum suppression,
or the observation of its decay into $\psi(2S) \, \gamma$ \cite{Babar:2008rn},
which disfavors $2^-$ because of multipole suppression.
\end{itemize}
Measurements of the mass of the $X(3872)$ in the 
$J/\psi \, \pi^+ \pi^-$ decay channel \cite{Jpsipipi}
imply that its energy relative to the $D^{*0} \bar D^0$ threshold is 
\beq
M_X - ( M_{D^{*0}} + M_{D^0} ) = -0.42 \pm 0.39 ~ {\rm MeV}.
\label{MX}
\eeq
The mass of the $X(3872)$ has also been measured in the 
$D^0 \bar D^0 \pi^0$ decay channel \cite{Gokhroo:2006bt}
and in the $D^{*0} \bar D^0$ decay channel \cite{DstarD}.
Measurements in the $D^0 \bar D^0 \pi^0$ channel are
biased towards larger values by a contribution from a 
threshold enhancement of $D^{*0} \bar D^0$ above the 
threshold \cite{Braaten:2007dw}. 
Measurements in the $D^{*0} \bar D^0$ channel are
further biased towards larger values by the analysis procedure 
that assigns an energy above threshold to $D^0 \bar D^0 \pi^0$
or $D^0 \bar D^0 \gamma$ events below the threshold \cite{Stapleton:2009ey}.
The quantum numbers $1^{++}$ imply that the $X(3872)$ has an S-wave 
coupling to $D^{*0} \bar D^0$.  Its tiny energy relative to the
$D^{*0} \bar D^0$ threshold implies that it is a resonant coupling.  
This system is therefore governed by the universal behavior of particles 
with short-range interactions and an S-wave  threshold resonance 
that is predicted by nonrelativistic quantum mechanics \cite{Braaten:2004rn}.
The universal properties of the system are determined
by the pair scattering length $a$ only.
If $a>0$, one of the universal properties is the binding energy:
$E_X = 1/(2 \mu a^2)$, where $\mu$ is the reduced mass of the pair.
Another universal property is the root-mean-square separation 
of the constituents: $r_X = a/\sqrt2$.  
Identifying the mass difference in Eq.~(\ref{MX}) as $-E_X$,
we find that the charm mesons in the $X(3872)$
have an astonishingly large rms separation: $r_X = 4.9^{+13.4}_{-1.4}$~fm. 

The universal properties of an S-wave threshold resonance
in the 2-body sector
can be derived from the universal transition amplitude for 
scattering of the constituents, which is a function of the
total energy $E$ of the pair relative to the scattering threshold 
in their center-of-momentum frame:
\beq
{\cal A}(E)  = \frac{2 \pi/\mu}{- 1/a + \sqrt{- 2 \mu E - i \epsilon}}.
\label{A-E}
\eeq
If $a$ is positive, 
this amplitude has a pole at the energy of the bound state:
$E = - E_X$, where $E_X = 1/(2 \mu a^2)$.
The rate for a process involving the bound state can 
be calculated diagrammatically by introducing a vertex for the 
coupling of the bound state to its constituents.  
Up to an irrelevant phase, the coupling constant $G$ for the vertex is the  
square root of the residue of the pole in the amplitude in  
Eq.~(\ref{A-E}) at $E = - E_X$:
\beq
G = 
\left( 8 \pi^2 E_X/\mu^3 \right)^{1/4}.
\label{GXDD}
\eeq
Taking into account 
the amplitude $1/\sqrt2$ for the constituents of $X$ to be  
$D^{*0} \bar D^0$ or $D^0 \bar D^{*0}$ from Eq.~(\ref{X-DD}),
the vertex factor for the coupling of $X$ to either pair of charm mesons 
is $G/\sqrt{2}$.

\section{Charm meson scattering into $\bm{X \pi}$}
\label{sec:breakup}

The masses of many different particles enter into the 
cross section that we will calculate.  It is therefore useful to introduce a
compact notation for these masses.
We denote the masses of $D^{*0}$ and $D^{*+}$ by $M_{*0}$ and $M_{*+}$,
or by $M_{*}$ if the difference can be neglected.
We denote the masses of $D^{0}$ and $D^{+}$ by $M_{0}$ and $M_{+}$,
or by $M$ if the difference can be neglected.
We denote the masses of $\pi^{0}$ and $\pi^{+}$ by $m_{0}$ and $m_{+}$,
or by $m_\pi$ if the difference can be neglected.
The mass of the $X(3872)$ is $M_X = M_{*0} + M_{0} - E_X$,
where $E_X$ is its binding energy.
The differences between the $D^*$ mass and the sum of the 
$D$ and $\pi$ masses are denoted by
\bqa
\delta_{00} &=& M_{*0} - M_{0} - m_{0} = 7.14 \pm 0.07 ~ {\rm MeV},
\label{delta000}
\\
\delta_{0+} &=& M_{*+} - M_{0} - m_{+}  = 5.85 \pm 0.01~ {\rm MeV}.
\label{delta101}
\eqa
We will refer to the energy scale set by these mass differences 
as $\delta$.

In XEFT, there is an extremely small energy scale:  
the binding energy of the $X(3872)$, which is less than 1~MeV.
In the generalization of XEFT that includes charged charm mesons 
and charged pions, there is also the relatively small energy scale
$\delta$.  The reason the $D^*$ mesons have relatively long
lifetimes is because $\delta$ is small compared to $m_\pi$,
which makes their hadronic decay rates comparable to 
their electromagnetic decay rates.
A pion with momentum comparable to $m_\pi$
is generally referred to as a {\it soft pion}.  
We will refer to a pion with kinetic energy comparable to 
or smaller than $\delta$ as an {\it ultrasoft pion}.  
The momentum scale for an ultrasoft pion
is $\sqrt{m_\pi \delta}$, which is about 30~MeV.

Two types of small mass ratios that appear in XEFT 
calculations are the ratio of a pion mass to a charm meson mass,
such as $m_0/M_0= 0.072$ and $m_0/M_{*0}= 0.067$,
and the ratio of a mass difference $\delta$ to a pion mass,
such as $\delta_{0+}/m_0 = 0.043$ and $\delta_{00}/m_0 = 0.053$. 
The numerical values of these ratios are all about $5\%$.
We therefore treat all the ratios $m_\pi/M_{(*)}$ and $\delta/m_\pi$ 
as order $\epsilon$.
The ratio of the binding energy of the $X(3872)$ to its mass is 
much smaller: $E_X/M_X = 0.0001$ for $E_X = 0.42$~MeV.
We will treat this ratio as $O(\epsilon^3)$.
We will organize our calculations to include all terms suppressed by  
$O(\epsilon)$, but we will feel free to drop corrections of $O(\epsilon^2)$.
Thus a multiplicative factor of the mass of $X(3872)$ can be approximated
as $M_X\approx M_{*0} + M_0$.
The reduced mass of $D^{*0}$ and $\bar D^{0}$ can be expressed as 
$\mu \approx M M_*/M_X$, since the corrections are $O(E_X/M_X)$ 
and hence $O(\epsilon^3)$ in our power counting.
Likewise, the reduced mass of $\pi^+$ and $X$ is
$\mu_{\pi^+ X} \approx M_X m_+/(2 M_*)$ and the reduced mass of 
$\pi^+$ and $D^0$ is $\mu_{\pi^+ D} \approx M m_+/M_*$. 
The corrections to these approximations are 
suppressed by $\delta/M_{(*)}$ and hence $O(\epsilon^2)$.

\begin{figure}[t]
\centerline{\includegraphics*[width=6cm,angle=0,clip=true]{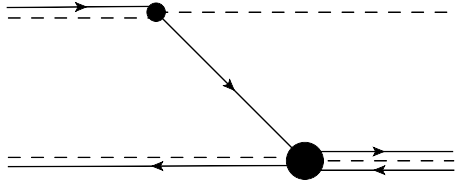}}
\vspace*{0.0cm}
\caption{
Feynman diagram for the reaction $D^{*+} \bar D^{*0} \to  X \pi^+$.
Spin-1 and spin-0 charm mesons are represented by solid+dashed and
solid lines, respectively.  Pions are represented by dashed lines.
The $X(3872)$ is represented by a solid+dashed+solid line.
}
\label{fig1}
\end{figure}

We would like to calculate the cross section for the scattering of
an ultrasoft pion and the $X(3872)$ into a pair of 
spin-1 charm mesons $D^* \bar D^*$.
We find it convenient to first calculate the cross section 
for the time-reversed process: the formation of $X(3872)$
by the reaction $D^* \bar D^* \to  X \pi$.
At leading order in the pion transitions,
this reaction proceeds only through the Feynman diagram 
in Fig.~\ref{fig1}.  Our notation for the lines representing 
the $X$, the charm mesons, and the pion is such that the number 
of solid lines and dashed lines is conserved at every vertex. 
The pion is ultrasoft if the {\it collision energy} of the 
$D^*$ and $\bar D^*$, which is their kinetic energy 
in the center-of-mass frame, is small compared to $m_\pi$.
In the diagram in Fig.~\ref{fig1}, the virtual $D$ 
is off its energy shell by an amount that is small compared to $m_\pi$.
Thus this reaction is within the region of validity of XEFT.
The reactions $D^* \bar D \to  X \pi$ and $D \bar D \to  X \pi$
can proceed through diagrams similar to the one in Fig.~\ref{fig1}
except that the number of dashed lines is not conserved at every vertex.
The virtual charm meson that is exchanged must therefore 
be off its energy shell by an amount of order $m_\pi$.
Thus these reactions can not be described by XEFT.

We proceed to calculate the cross section for the reaction 
$D^{*+} \bar D^{*0} \to  X \pi^+$.  We take the incoming momenta 
of the $D^{*+}$ and $\bar D^{*0}$ to be $\pm \bm{p}$, 
and we take the outgoing momenta of the $\pi^+$ and $X$ to be
$\pm \bm{k}$.  We take the collision energy 
$E_{\rm cm} = p^2/M_*$ to be small compared to $m_\pi$, 
so the $\pi^+$ is ultrasoft.  The T-matrix element is
\beq
{\cal T} = 
\frac{(G/\sqrt{2}) (g/f_\pi) (\bm{k} \cdot \bm{\epsilon}_{+})
                                  (\bm{\epsilon}_X^* \cdot \bm{\epsilon}_{0})}
     {E_X + [\bm{p} - (M_*/M_X) \bm{k}]^2/(2 \mu)},
\label{TpiX}
\eeq
where $\bm{\epsilon}_X^*$, $\bm{\epsilon}_{+}$, and $\bm{\epsilon}_{0}$ are 
the polarization vectors of the $X$, $D^{*+}$, and $\bar D^{*0}$,
respectively.  The $D^*D\pi$ coupling constant $g/f_\pi$
can be determined from the measured partial width for 
$D^{*+} \to D^{0} \pi^+$:  
\beq
g/f_\pi = 2.8 \times 10^{-4}~{\rm MeV}^{-3/2}.
\label{gf}
\eeq
The coupling constant $g$ defined here differs from the dimensionless 
coupling constant $g=0.6$ used in Refs.~\cite{Fleming:2007rp,Fleming:2008yn} 
by a factor of $\sqrt{2 m_\pi}$.
The simple expression in the denominator of Eq.~(\ref{TpiX})
for the propagator of the
virtual $D^0$ can be obtained most easily by exploiting the approximate 
Galilean invariance of this reaction.
Galilean invariance holds when the sum of the masses is exactly 
equal in the initial and final states.
For the reaction $D^{*+} \bar D^{*0} \to X \pi^+$, the sum of the masses 
decreases between the initial and  final states by only about
$\delta_{0+} \approx 5.85$~{\rm MeV}, which is about 1 part in 700.
We can take advantage of the approximate Galilean invariance 
by calculating the $D$ propagator in the rest frame of the $X(3872)$.
The momentum $\bm{p} - (M_*/M_X) \bm{k}$ in the denominator
of Eq.~(\ref{TpiX}) is just the momentum transferred through 
the virtual $D$ meson in that frame. 
The propagator calculated in the center-of-mass frame 
involves the sum or difference of three kinetic energies.  
Those three terms can be combined to give the single kinetic energy 
in the denominator of Eq.~(\ref{TpiX}) plus terms that have been 
neglected because they are suppressed by a factor of 
$E_X/M_X = O(\epsilon^3)$.

Squaring the T-matrix element in Eq.~(\ref{TpiX})
and averaging and summing over the spins of the 
$D^{*+}$, $\bar D^{*0}$, and $X$, we get
\beq
\frac{1}{9} \sum_{\rm spins}| {\cal T} |^2 = 
\frac{2 G^2 (g/f_\pi)^2 \mu^2 k^2}
     {3 \left[ 2 \mu E_X + p^2 - 2 (M_*/M_X)(\bm{p} \cdot \bm{k}) 
                         + (M_*/M_X)^2 k^2 \right]^2}.
\label{TsqpiX}
\eeq
After integrating over the momenta of the outgoing $X$ and $\pi$, 
our final result for the reaction rate is
\beq
v_{\rm rel} \sigma [D^{*+} \bar D^{*0} \to  X \pi^+] = 
\frac{2 G^2 (g/f_\pi)^2 \mu^2 \mu_{\pi X} k^3}
     {3 \pi \Delta(p^2,(M_*^2/M_X^2) k^2, -2 \mu E_X)},
\label{sigDD}
\eeq
where $\Delta$ is the triangle function:
$\Delta(a,b,c) = a^2 + b^2 + c^2 -2(ab+bc+ca)$.
The pion momentum $k$ is determined by conservation of energy:
\beq
\frac{k^2}{2 \mu_{\pi X}} =
\delta_{0+}  + E_X + \frac{p^2}{M_*}.
\label{Econ}
\eeq
By completing the square in the variable $p^2$ in the denominator
of Eq.~(\ref{sigDD}) and dividing by the relative velocity 
$v_{\rm rel} = 2 p/M_*$, we obtain our final expression 
for the cross section:
\beq
\sigma [D^{*+} \bar D^{*0} \to  X \pi^+] = 
\frac{G^2 (g/f_\pi)^2 M_X M_*^2 m_\pi k^3/(24 \pi p)}
     {[p^2 - (M_* /2 M)(m_\pi \delta_{0+} - M_X E_X )]^2 
      + M_*^2 m_\pi \delta_{0+} M_X E_X/M^2}.
\label{sigDD-1}
\eeq
At small momentum $p$, the cross section diverges as $1/p$ but the 
reaction rate $v_{\rm rel} \sigma [D^{*+} \bar D^{*0} \to  X \pi^+]$
is well-behaved.  The asymptotic behavior of the cross section 
for large momentum $p$ is
\beq
\sigma [D^{*+} \bar D^{*0} \to  X \pi^+] \longrightarrow
\frac{(g/f_\pi)^2 M_X M (m_\pi/\mu)^{5/2} (2 M_X E_X)^{1/2}}{12 p^2}.
\label{kmax}
\eeq
%

\begin{figure}[t]
\centerline{\includegraphics*[width=12cm,angle=0,clip=true]{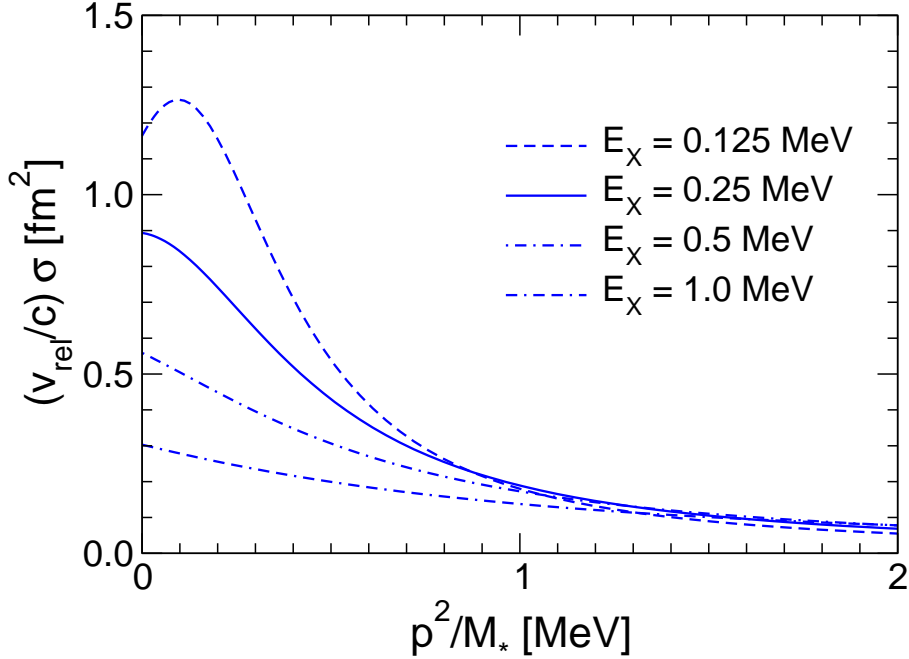}}
\vspace*{0.0cm}
\caption{
Reaction rate $v_{\rm rel} \sigma$ for $D^{*+} \bar D^{*0} \to  X \pi^+$
as a function of the collision energy $E_{\rm cm} = p^2/M_*$ 
for several values of the binding energy $E_X$.
}
\label{fig2}
\end{figure}

In Fig.~\ref{fig2}, we show the reaction rate 
$v_{\rm rel} \sigma [D^{*+} \bar D^{*0} \to  X \pi^+]$  
as a function of the 
$D^{*+} \bar D^{*0}$ collision energy $E_{\rm cm} = p^2/M_*$ for 
several values of the binding energy $E_X$ of the $X(3872)$.  From 
the expression in Eq.~(\ref{sigDD-1}), we can see that
there are two competing momentum scales that govern the behavior
of the reaction rate at small $p$:
$(m_\pi \delta_{0+})^{1/2} \approx 30$~MeV and $(M_X E_X)^{1/2}$,
which is approximately 60~MeV if $E_X=1$~MeV.
If the binding energy $E_X$ decreases to less than about 
$m_\pi \delta_{0+}/M_X \approx 0.2$~MeV, 
the peak in the reaction rate shifts from zero collision energy
to a positive value near $(m_\pi \delta_{0+} - M_X E_X )/(2 \mu)$.
This is the collision energy for which there is no momentum transferred 
through the virtual $D^0$. The peak at nonzero collision energy
is evident in the curve for $E_X = 0.125$~MeV in Fig.~\ref{fig2}.

The reaction rate for $D^{*0} \bar D^{*0} \to  X \pi^0$
can be calculated in a similar way.
In addition to the Feynman diagram in Fig.~\ref{fig1},
which involves exchange of a virtual $D^0$, 
there is a second diagram that involves exchange 
of a virtual $\bar D^0$. 
The T-matrix element is therefore the sum of two terms.
The term corresponding to the Feynman diagram in Fig.~\ref{fig1} 
differs from the expression in Eq.~(\ref{TpiX}) by 
an isospin Clebsch-Gordan factor of $1/\sqrt{2}$.
In the energy conservation condition in Eq.~(\ref{Econ}),
$\delta_{0+}$ must be replaced by $\delta_{00}$. 

We now consider the reaction 
$\pi^+ X \to D^{*+} \bar D^{*0}$.  We take the incoming momenta 
of the $ \pi^+$ and $X$ to be $k$ and we take the outgoing momenta 
of the $D^{*+}$ and $\bar D^{*0}$ to be $p$. 
They are related by the conservation of energy condition in 
Eq.~(\ref{Econ}).  The collision energy is $E_{\rm cm} = k^2/(2 \mu_{\pi X})$.
The energy threshold for the breakup process is $\delta_{0+} + E_X$.
The cross section can be obtained from the 
production cross section in Eq.~(\ref{sigDD-1}) by changing the 
flux factor from $M_*/(2 p)$ to $\mu_{\pi X}/k$ and by 
changing the phase space factor from $\mu_{\pi X} k/\pi$ 
to $M_* p/(2 \pi)$: 
\beq
\sigma [X \pi^+ \to D^{*+} \bar D^{*0}] = 
\frac{G^2 (g/f_\pi)^2 M_X M_*^2 m_\pi k p/(24 \pi)}
     {[p^2 - (M_* /2 M)(m_\pi \delta_{0+} - M_X E_X )]^2 
      + M_*^2 m_\pi \delta_{0+} M_X E_X/M^2}.
\label{sigXpiDD}
\eeq
%

\section{$\bm{\pi X}$ Elastic scattering}
\label{sec:elastic}

\begin{figure}[t]
\centerline{\includegraphics*[width=8cm,angle=0,clip=true]{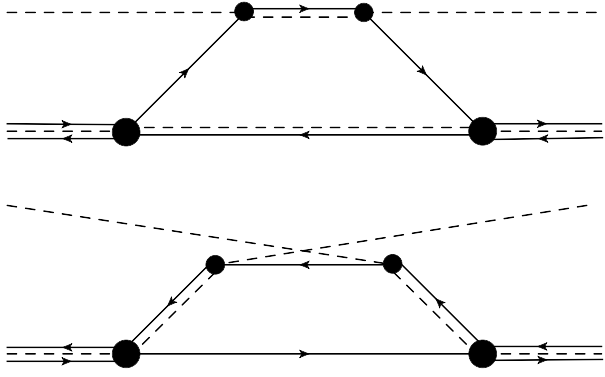}}
\vspace*{0.0cm}
\caption{
Two of the six one-loop Feynman diagrams for $\pi^+ X$ elastic scattering.  
The other four diagrams involve a Weinberg-Tomazawa 
vertex and add up to zero.
}
\label{fig3}
\end{figure}

Another process in the $D \bar D \pi \pi$ sector 
that is calculable using XEFT is $\pi^+ X$ elastic scattering.
At leading order in the pion transitions, 
$\pi^+ X$ elastic scattering proceeds through six one-loop diagrams.
The two diagrams in Fig.~\ref{fig3} involve
virtual $D^{*+}$ and $\bar D^{*0}$ mesons and virtual 
$D^0$ and $D^-$ mesons, respectively.
There are also four additional diagrams with 
Weinberg-Tomozawa vertices.  One of them is the 
first diagram in Fig.~\ref{fig3}
with the virtual $D^{*+}$ propagator shrunk to a point
to obtain a $\pi^+ D^0 - \pi^+ D^0$ vertex.
Another one is obtained from the second diagram by shrinking
the virtual $D^{-}$ propagator
to a point to obtain a $\pi^+ \bar D^{*0} - \pi^+ \bar D^{*0}$ vertex.
The other two diagrams involve a $\pi^+ D^{*0} - \pi^+ D^{*0}$ vertex
and a $\pi^+ \bar D^0 - \pi^+ \bar D^0$ vertex, respectively.

In Ref.~\cite{Fleming:2007rp}, the authors carried out a 
power-counting analysis for XEFT that determined that the
$D^* - D \pi$ vertices can be treated perturbatively.  
They did not consider the Weinberg-Tomozawa (WT) vertices 
for $D \pi - D \pi$ and $D^* \pi - D^* \pi$. These come from the  
kinetic term in the HH$\chi$PT Lagrangian,
\bea\label{hhchipt}
{\cal L} &=& {\rm Tr}[H^\dagger_a (i D_0)_{ba} H_b] \, ,
\eea
where the chiral covariant derivative is 
\bea\label{cov}
(D_0)_{ba} =
\partial_0 \delta_{ba} -\frac{1}{2f^2}[\pi,\partial_0 \pi]_{ba} + O(\pi^4)\, ,
\eea
and our definitions for the fields $H_b$, $H^\dagger_a$, 
and $\pi$ can be found in Appendix A of Ref.~\cite{Fleming:2007rp}.
The WT coupling for $D^0$ mesons is  obtained by plugging 
the second term in Eq.~(\ref{cov})  into Eq.~(\ref{hhchipt}), yielding
\bea\label{WTr}
{\cal L}_{\rm WT} = \frac{1}{2 f^2}
P_0^\dagger \pi^- i \overleftrightarrow{\partial_0}  \pi^+ P_0 \, ,
\eea
where $P_0$ is the field for  the $D^0$. There is a similar coupling 
for the $D^{*0}$. 
Next we rewrite Eq.~(\ref{WTr}) in terms of nonrelativistic pion fields, 
as is appropriate for the XEFT Lagrangian. The relativistic fields 
$\pi^\pm$ when written in terms of nonrelativistic fields  are
\bea
\pi^\pm = \frac{1}{\sqrt{2 m_\pi}}
\left( e^{- i m_\pi t}\hat{\pi}^\pm 
+ e^{i m_\pi t}\hat{\pi}^{\mp \dagger} \right),
\eea
where the $\hat{\pi}^\pm \,(\hat{\pi}^{\pm \dagger})$ denotes 
a nonrelativistic field that annihilates (creates) $\pi^\pm$ mesons. 
The XEFT Lagrangian for the WT vertex is 
\bea\label{WTnr}
{\cal L}_{\rm WT} = \frac{1}{2 f^2}P_0^\dagger 
(\hat{\pi}^{+\,\dagger} \hat\pi^+ -\hat{\pi}^{- \dagger}\hat\pi^-)  P_0 \,,
\eea
with a similar term for $D^{0*}$ mesons.
We have kept only the terms in which the phase factors 
$e^{\pm i m_\pi t}$ cancel. Other terms describe processes 
that are outside the range of validity of XEFT.  
The  WT vertex is $O(Q^0)$ in the power counting of Ref.~\cite{Fleming:2007rp}.
This means a diagram with a WT vertex is the same order as a diagram 
obtained by replacing the WT vertex 
with  a virtual $D$ or $D^*$ propagator and two $D^*-D\pi$ couplings,
since $D^* - D \pi$ vertices are  $O(Q)$ and $D$ and $D^*$ propagators 
are  $O(1/Q^2)$.  The one-loop diagrams for $\pi^+ X(3872)$  
elastic scattering with a WT vertex are then 
the same order in $Q$ as the diagrams in Fig.~\ref{fig3}. 
However, applying charge conjugation to Eq.~(\ref{WTnr}),
we see that the Feynman rule for $\pi^+ D^{0(*)} \to \pi^+ D^{0(*)}$ 
has the opposite sign as the Feynman rule
for $\pi^+ \bar{D}^{0(*)} \to \pi^+ \bar{D}^{0(*)}$, 
so the four  diagrams with a WT vertex add up to zero.  
The only nonvanishing one-loop contribution to $\pi^+ X(3872)$ 
elastic scattering comes from the diagrams in Fig.~\ref{fig3}.

An explicit one-loop calculation of $\pi^+ X$ 
elastic scattering would be worthwhile as a test of the 
calculational technology of XEFT.
For collision energies below the break-up threshold
$\delta_{0+} + E_X$, the scattering amplitude will be real-valued.
As the binding energy $E_X$ decreases to 0, 
the constituents of the $X(3872)$ have a larger 
and larger probability of being well separated. 
The cross section near threshold should therefore reduce 
in this limit to the sum of the cross sections 
for scattering off the individual constituents.

\section{Summary}
\label{sec:summary}

XEFT is an effective field theory that was originally designed
to describe systems consisting of $D^{*0} \bar D^0$, 
$D^0 \bar D^{*0}$, and $D^0 \bar D^0 \pi^0$ with total energy
relative to the $D^{*0} \bar D^0$ threshold
that is small compared to the 8~MeV $D^{*+} D^-$ threshold.
The only important energy scale in this effective field theory 
is the binding energy $E_X$ of the $X(3872)$.
XEFT can be generalized to an effective field theory that 
includes charged charm mesons and charged pions
and describes systems 
consisting of $D^* \bar D$, $D \bar D^{*}$, 
and $D \bar D \pi$ with total energy 
relative to the $D^* \bar D$ threshold that is small 
compared to $m_\pi$.
In addition to the tiny energy scale $E_X$, 
this effective field theory also describes the ultrasoft 
energy scale $\delta$ set by the difference between 
$D^*-D$ mass splittings and $m_\pi$. 
We have pointed out that XEFT can also be applied to systems 
consisting of $D^* \bar D^*$, $D^* \bar D \pi$, $D \bar D^{*} \pi$, 
and $D \bar D \pi \pi$ with total energy 
relative to the $D^* \bar D^*$ threshold that is small 
compared to $m_\pi$.
We have used XEFT to calculate the cross section 
for $\pi^+ X \to D^{*+} \bar D^{*0}$ for ultrasoft collision energy.
This cross section is completely determined by the masses
and widths of the charm mesons, the pion masses, 
and the binding energy of the $X(3872)$.

\begin{acknowledgments}
This research was supported in part by the Department of Energy 
under grant DE-FG02-05ER15715, by the Alexander von Humboldt Foundation,
by the DFG through SFB/TR 16 \lq\lq Subnuclear structure of matter'', 
and by the BMBF under contract No. 06BN9006. 
We express our appreciation to the Physikzentrum in Bad Honnef, 
Germany, where this project was initiated, and to the 
Institute for Nuclear Theory in Seattle, 
where part of this work was carried out.
\end{acknowledgments}

\end{document}